\documentclass[letterpaper,twocolumn,prl,
aps,showpacs,groupedaddress,floatfix]{revtex4-1}
\usepackage[latin1]{inputenc}
\usepackage{bm}
\usepackage[usenames]{color}
\usepackage{multirow}
\usepackage{amssymb}
\usepackage{amsbsy}
\usepackage{mathtools}
\usepackage{amsmath}
\usepackage{stmaryrd}
\usepackage{graphicx}
\usepackage{epsfig}
\usepackage{placeins}
\usepackage{ulem}
\usepackage{bbold}
\usepackage{braket}
\usepackage{blindtext}
\usepackage[colorlinks,linkcolor=blue,citecolor=blue,urlcolor=blue]{hyperref}

\usepackage{tikz}

\newcommand{\UCSymbol}[1]{%
\begin{tikzpicture}[#1]%
\filldraw (0,5pt) circle (1.2pt);
\draw[line width = 0.5pt] (-0.1,5pt) -- (0.1,5pt);%
\end{tikzpicture}%
}
\usepackage{scalerel}
\usetikzlibrary{calc}
\usetikzlibrary{patterns}
\usetikzlibrary{svg.path}
\definecolor{orcidlogocol}{HTML}{A6CE39}
\tikzset{
  orcidlogo/.pic={
    \fill[orcidlogocol] svg{M256,128c0,70.7-57.3,128-128,128C57.3,256,0,
    198.7,0,128C0,57.3,57.3,0,128,0C198.7,0,256,57.3,256,128z};
    \fill[white] svg{M86.3,186.2H70.9V79.1h15.4v48.4V186.2z}
                 svg{M108.9,79.1h41.6c39.6,0,57,28.3,57,53.6c0,27.5-21.5,
                 53.6-56.8,53.6h-41.8V79.1z 
M124.3,172.4h24.5c34.9,0,42.9-26.5,
42.9-39.7c0-21.5-13.7-39.7-43.7-39.7h-23.7V172.4z}
                 svg{M88.7,56.8c0,5.5-4.5,10.1-10.1,10.1c-5.6,
0-10.1-4.6-10.1-10.1c0-5.6,4.5-10.1,10.1-10.1C84.2,46.7,88.7,51.3,88.7,56.8z};
  }
}

\newcommand\orcid[1]{\href{https://orcid.org/#1}{\mbox{\scalerel*{
\begin{tikzpicture}[yscale=-1,transform shape]
\pic{orcidlogo};
\end{tikzpicture}
}{|}}}}

\makeatletter

\begin{document}
\title{Many-body localization and delocalization dynamics in 
the thermodynamic limit}

\author{Jonas Richter~\hspace{-0.1cm}\orcid{0000-0003-2184-5275}}
\email{j.richter@ucl.ac.uk}
\affiliation{Department of Physics and Astronomy, University College London, 
Gower Street, London WC1E 6BT, UK}

\author{Arijeet Pal~\hspace{-0.1cm}\orcid{0000-0001-8540-0748}}
\affiliation{Department of Physics and Astronomy, University College London, 
Gower Street, London WC1E 6BT, UK}

\date{\today}

%---------------------------------------------------------------------------------------------------
\begin{abstract}

Disordered quantum systems undergoing a many-body 
localization (MBL) transition fail to reach thermal equilibrium under 
their own dynamics. Distinguishing between asymptotically localized or  
delocalized dynamics
based on numerical results is however nontrivial due to finite-size  
effects. Numerical linked cluster expansions (NLCE) provide a means 
to tackle quantum systems directly in the thermodynamic limit, 
but are challenging for models without translational invariance. 
Here, we demonstrate that NLCE provide a powerful tool to explore 
MBL by simulating quench dynamics in disordered spin-$1/2$ two-leg  
ladders and Fermi-Hubbard chains.
Combining NLCE with an efficient real-time evolution of pure states, we obtain 
converged results for the decay of the imbalance on long time scales and show 
that, especially for intermediate disorder below the putative MBL 
transition, NLCE outperform direct simulations of finite systems 
with open or periodic boundaries. Furthermore, while spin is 
delocalized even in strongly disordered 
Hubbard 
chains with frozen charge, we unveil that an additional tilted potential leads 
to 
a drastic slowdown of the spin imbalance and nonergodic behavior on 
accessible times. Our 
work sheds light on MBL in systems beyond the well-studied disordered 
Heisenberg chain and emphasizes the 
usefulness of NLCE for this purpose.  
  
\end{abstract}

\maketitle
%-------------------------------------------------------------------------------
%--------------------

{\it Introduction.--}
Many-body localization (MBL) extends Anderson localization to interacting 
quantum systems~\cite{Nandkishore2015, Abanin2019}. Based on seminal early 
works 
\cite{Gornyi2005, Basko2006}, and numerous subsequent studies (see e.g.\ 
\cite{Oganesyan2007, Pal2010, Imbrie2016, 
Berkelbach2010, Luitz2015, Kjall2014, Bera2015}), it is believed that 
disordered one-dimensional 
(1d)
system with local interactions can undergo a 
transition from a thermal phase to a MBL phase for sufficiently strong 
disorder. The MBL phase is characterized, e.g., by a breakdown of the 
eigenstate thermalization hypothesis \cite{Dallesio2016}, area-law entangled 
energy 
eigenstates \cite{Bauer2013}, and a logarithmic growth of entanglement in 
time \cite{Znidaric2008, Bardarson2012}. 
Its properties can be 
understood in terms of an emergent set of local integrals of motion 
\cite{Serbyn2013, Huse2014, Chandran2015, Ros2015}, 
so-called l-bits.
Due to a finite overlap with these 
l-bits, observables fail to thermalize under time 
evolution, which makes 
MBL systems candidates for realizing quantum memories. This memory of 
initial conditions is a key experimental signature of MBL \cite{Schreiber2015, 
Choi2016, Smith2016}, but is theoretically investigated as well \cite{Luitz2017, 
Enss2017, Nico-Katz2021}.

The emergent l-bit phenomenology of MBL motivated by nearest-neighbor qubit 
models can become unstable in higher dimensions \cite{Choi2016, Wahl2019, 
Decker2021, Chertkov2021}, in the presence of non-Abelian symmetries 
\cite{parameswaran2018many, potter2016symmetry, protopopov2017effect, 
protopopov2020non}, long-range interactions \cite{Yao2014, Burin2015, 
Nandkishore2017, Modak2020}, large local Hilbert-space dimensions 
\cite{Richter2019_2, 
Richter2020_3, Schliemann2021}, and disorder-free systems \cite{Grover2014, 
Schiulaz2014, Brenes2018, Heitmann2020, 
Smith2017, Yao2016, Sirker2019, Schulz2019, vanNieuwenburg2019, Karpov2021}. 
Despite of these instabilities of the fully many-body localized systems, they 
can show anomalously slow dynamics and even nonergodic behavior for certain 
initial conditions \cite{Gopalakrishnan2020}, referred to as MBL 
\textit{regime} \cite{Morningstar2021}. For instance, in two dimensions (2d) 
signatures of the MBL regime exist in experiments and numerics \cite{Choi2016, 
Wahl2019, Decker2021, Chertkov2021}, although in the thermodynamic limit the 
avalanche picture \cite{DeRoeck2017, Doggen2020} suggests fully chaotic 
dynamics, albeit at astronomically long time scales. 
\begin{figure}[tb]
 \centering
 \includegraphics[width=0.85\columnwidth]{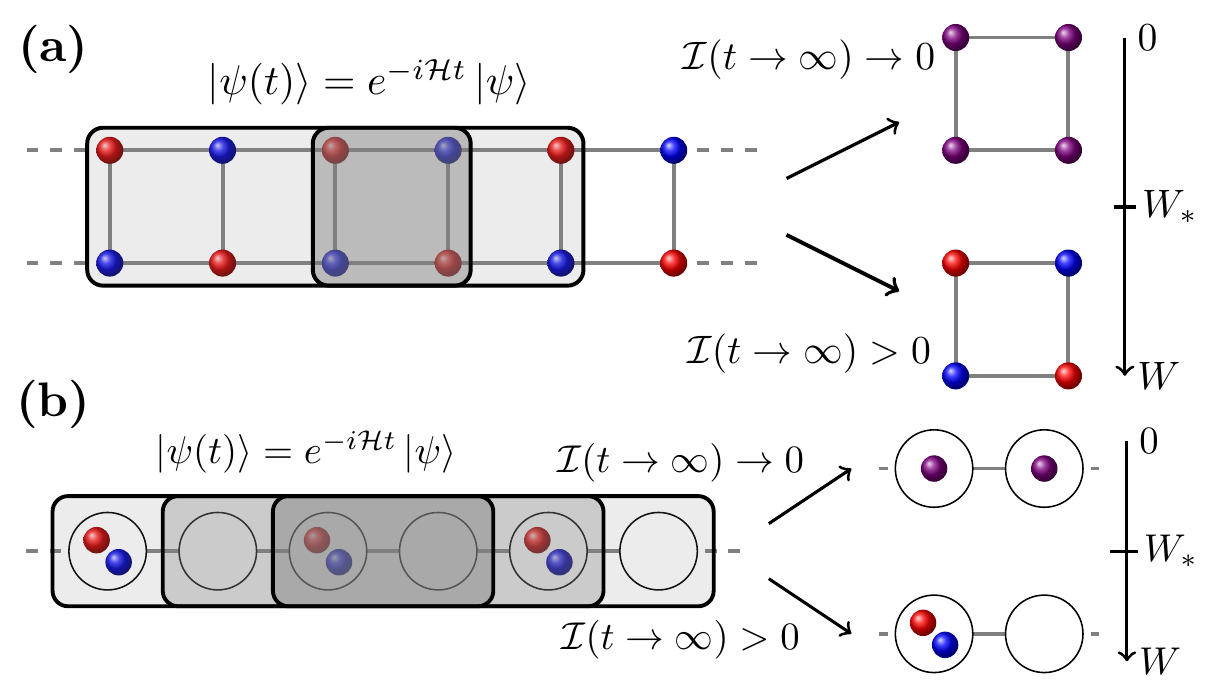}
 \caption{We study the imbalance ${\cal I}(t)$ in disordered  
{\bf (a)} two-leg spin ladders and {\bf (b)} 
Fermi-Hubbard chains. (Disorder not shown here.) For $W < 
W_\ast$, 
${\cal I}(t)$ is expected to decay to zero, while ${\cal I}(t) > 0$ in the 
MBL phase for $W > W_\ast$. NLCE is used to simulate ${\cal I}(t)$ in the 
thermodynamic limit $L \to \infty$.
Within NLCE, ${\cal I}(t)$ is obtained on finite clusters
(shaded rectangles), whose contributions are 
 suitably combined to yield quantum dynamics without finite-size effects 
\cite{SuppMat}.}
 \label{fig::1}
\end{figure}

The main complication for numerical studies of MBL is the presence of strong 
finite-size effects \cite{Weiner2019, Panda2020, Sierant2021, Doggen2018, 
Abanin2021, Sierant2020, Morningstar2021}.
In this context, the existence of a genuine MBL phase (even for 
nearest-neighbor 
1d models) has been put into question \cite{Suntajs2020, 
Kiefer-Emmanouilidis2020, Sels2021}. 
Providing a definite answer to this issue by means of numerical approaches 
is challenging.
On one hand, full or sparse-matrix 
diagonalization methods are restricted to 
intermediate system 
sizes, potentially leading to inconclusive results.  
On the other hand, tensor-network techniques can treat 
large systems, but the times reachable in simulations 
are limited by the growth of entanglement \cite{Paeckel2019}.
Despite notable progress to extend these time 
scales \cite{Doggen2018, Doggen2020}, and the development of other 
sophisticated  
methods \cite{Burau2021, Thomson2018, Kvorning2021}, studying  
quantum many-body dynamics, especially beyond 
1d, remains difficult \cite{Kennes2018, Hubig2019, Kshetrimayum2020}.

In this Letter, we study the nature of the MBL regime in two classes of 
disordered models (see Fig.\ \ref{fig::1}), (i) spin-$1/2$ two-leg ladders 
\cite{Doggen2020, Baygan2015, Wiater2018, Hauschild2016}, a quasi 
one-dimensional system which represents an intermediate case between a 1d chain 
and a 2d lattice, and (ii) Fermi-Hubbard (FH) chains  \cite{Prelovsek2016, 
Mondaini2015, Kozarzewski2018, Zakrzewski2018, Iadecola2019, Protopopov2019, 
Kurlov2021}, where disorder only couples to the 
charge degrees of freedom. Both of them can also be viewed as 
1d models with local Hilbert-space 
dimension greater than two. In the FH chain, 
there is a SU$(2)$ symmetry incompatible with 
MBL, and 
we also study the effect 
of a tilted potential which can 
induce Stark MBL \cite{Schulz2019, vanNieuwenburg2019, Guo2021}. 

We demonstrate that numerical linked cluster expansions 
(NLCE) \cite{Tang2013} provide a powerful means to study the MBL regime. 
The crucial advantage of NLCE is that, if converged, they yield results 
directly in the thermodynamic limit, i.e., there are no finite-size effects.
We use NLCE to study the dynamics resulting from out-of-equilibrium
initial states and obtain 
converged results for the imbalance ${\cal I}(t)$ on long time 
scales, outperforming direct simulations of finite systems 
with open or periodic boundaries especially for intermediate disorder $W < 
W_\ast$, which allows the extraction of more 
accurate lower bounds for $W_\ast$. 
Furthemore, we show that, in contrast to strongly 
disordered FH chains where spin thermalizes despite charge being 
localized, an additional tilted 
potential leads to a slowdown of the spin imbalance and nonergodic 
behavior for certain initial states.

{\it Models \& Observables.--}
The first class of models we consider are disordered Heisenberg 
two-leg spin ladders, 
\begin{equation}\label{eq::Ham_Ladder}
 {\cal H}_\text{SL} = \sum_{k=1}^2\sum_{\ell} \left( {\bf S}_{\ell,k} 
{\bf S}_{\ell+1,k} + h_{\ell,k} S_{\ell,k}^z \right) + \sum_{\ell = 1}^{L} {\bf 
S}_{\ell,1} \cdot 
{\bf S}_{\ell,2}\ , 
\end{equation}
where ${\bf S}_{\ell,k} = (S_{\ell,k}^x,S_{\ell,k}^y,S_{\ell,k}^z)$ are 
spin-$1/2$ operators on leg $k$ and rung $\ell$, $L$ denotes the length of the 
ladder ($2L$ lattice sites in total), and the on-site fields 
$h_{\ell,k}\in[-W,W]$ are randomly drawn from a 
uniform distribution with $W$ setting the strength of disorder. We study the  
nonequilibrium dynamics resulting from quenches with antiferromagnetic initial 
states of the form [cf.\ Fig.\ \ref{fig::1}~(a)],
\begin{equation}\label{Eq::Psi0_Lad}
 \ket{\psi(0)} =  \left\vert \begin{array}{cccccccc}
\cdots & \downarrow & \uparrow & \downarrow & \uparrow & \downarrow & \uparrow 
& 
\cdots \\
\cdots & \uparrow & \downarrow & \uparrow & \downarrow & \uparrow & \downarrow 
& \cdots \\
\end{array}\right\rangle\ , 
\end{equation}
in the $\sum_{k,\ell} S_{\ell,k}^z = 0$ sector. 
We monitor the imbalance,
${\cal I}(t) = \sum_{k} \sum_{\ell} (-1)^{k+\ell} 
 \langle S_{\ell,k}^z(t)\rangle/L$,
where $\langle \cdot(t) \rangle = \bra{\psi(t)}\cdot \ket{\psi(t)}$, 
$\ket{\psi(t)} = e^{-i{\cal H}t}\ket{\psi(0)}$, and 
${\cal I}(0) = 1$. 
In case of thermalization, one expects 
$\lim_{t \to \infty} \lim_{L\to\infty} {\cal I}(t) \to 0$. In contrast, ${\cal 
I}(t)>0$ in the case of MBL, see Fig.\ \ref{fig::1}. Distinguishing between 
asymptotically localized or delocalized dynamics is challenging due to (i) 
finite-size effects and (ii) finite simulation times. In this Letter, we show 
that NLCE 
provide a means to mitigate the impact of (i) by obtaining ${\cal I}(t)$ in the 
thermodynamic limit $L \to \infty$. 

As a second model, we study disordered FH chains, 
\begin{equation}\label{Eq::Ham::Hub}
 {\cal H}_\text{FH} = 
-\hspace{-0.1cm}\sum_{\ell,\sigma}(c_{\ell,\sigma}^\dagger 
 c_{\ell+1,\sigma}^{} \hspace{-0.1cm}+ 
\text{h.c.}) +\hspace{-0.05cm} \sum_{\ell=1}^L( Un_{\ell,\uparrow}^{} 
n_{\ell,\downarrow}^{} + 
\mu_\ell^{} n_\ell^{})\ , 
\end{equation}
where $c_{\ell,\sigma}^\dagger$ ($c_{\ell,\sigma}^{}$) creates (annihilates) a 
fermion of spin $\sigma$ at site $\ell$, $U$ is the 
on-site interaction, $n_{\ell,\sigma}^{} = c_{\ell,\sigma}^\dagger 
c_{\ell,\sigma}^{}$, $n_\ell^{} = n_{\ell,\uparrow}^{} + 
n_{\ell,\downarrow}^{}$, 
and $\mu_\ell^{} = \epsilon_\ell^{} + V\ell$ with 
$\epsilon_\ell^{} 
\in [-W,W]$ is the spin-independent disorder with added 
tilt $V$ \cite{Schulz2019, vanNieuwenburg2019, Scherg2021, 
Guardado-Sanchez2020, Yao2021, Desaules2021}.
In our implementation, we exploit that ${\cal H}_\text{FH}$ can be 
mapped to a spin ladder, where the interactions are mediated by the rungs 
of the ladder \cite{Heitmann2020, Prosen2012}.   

We consider two experimentally relevant initial states \cite{
Scherg2018, Scherg2021}, i.e., density waves at half filling [cf.\ Fig.\ 
\ref{fig::1}~(b)],  
\begin{equation}\label{Eq::InitState_Hub}
 \ket{\psi_1(0)} = \prod_{\ell} c_{2\ell,\uparrow}^\dagger 
c_{2\ell,\downarrow}^\dagger \ket{0} =  \left\vert \begin{array}{cccccc}
\cdots\hspace{-0.1cm} & \uparrow \hspace{-0.05cm}\downarrow & 0 & 
\uparrow\hspace{-0.05cm}\downarrow \hspace{-0.1cm} & 0 &  
\cdots 
\end{array}\right\rangle\ , 
\end{equation}
or at quarter filling \cite{Mondaini2015}, both at zero magnetization, 
\begin{equation}\label{Eq::Init::Hub2}
  \ket{\psi_2(0)} = \prod_\ell c_{4\ell,\uparrow}^\dagger 
c_{4\ell+2,\downarrow}^\dagger \ket{0} = \left\vert \begin{array}{cccccc}
\cdots\hspace{-0.1cm} & \uparrow & 0 & \downarrow & 0\hspace{-0.1cm}  & 
\cdots 
\end{array}\right\rangle\ . 
\end{equation}
We simulate the charge and spin imbalances, 
${\cal I}_\text{ch}(t) \propto \sum_{\ell} \langle n_\ell(t) \rangle\langle 
n_\ell(0) \rangle$ 
and 
${\cal I}_\text{s}(t) \propto \sum_{\ell} \langle m_\ell(t)\rangle \langle 
m_\ell(0)\rangle$, with $m_\ell = n_{\ell,\uparrow} 
- 
n_{\ell,\downarrow}$, and ${\cal I}_\text{ch(s)}(0) = 
1$ 
\cite{NoteInitState}. 

While we are mainly interested in ${\cal I}(t)$ in the 
thermodynamic limit $L \to \infty$ using NLCE, we also consider finite systems 
with periodic boundary conditions (PBC) or open boundary conditions (OBC). For 
PBC, the first sums in Eqs.\ \eqref{eq::Ham_Ladder} and \eqref{Eq::Ham::Hub} run 
from $\ell = 1$ to $\ell = L$, with ${\bf S}_{L+1,k} = {\bf S}_{1,k}$,  
$c_{L+1,\sigma}^{(\dagger)} = c_{1,\sigma}^{(\dagger)}$, while in case of 
OBC they run up to $\ell = L-1$. As explained below, systems with 
OBC are a main ingredient within the NLCE formalism.

{\it Numerical linked cluster expansions.--}
NLCE provide a means to study 
quantum systems directly in the thermodynamic limit $L 
\to \infty$. 
The main idea is to write the quantity of interest as a 
sum over contributions from all clusters that can 
be embedded on the lattice \cite{Tang2013, Dusuel2010}. Originally 
introduced in the context of thermodynamics \cite{Rigol2006}, NLCE have also 
been used to study open quantum systems \cite{Biella2018}, entanglement 
entropies \cite{Kallin2013},
dynamical correlation functions \cite{Richter2019, Richter2020, Heitmann2021}, 
and quantum 
quenches in 1d and 2d systems \cite{Wouters2014, White2017, Mallayya2017, 
Mallayya2018, Guardado-Sanchez2018, Richter2020_2}. While NLCE are 
usually formulated for translational invariant systems, disordered systems can 
be treated as well \cite{Devakul2015, Mulanix2019, Park2021, Tang2015, 
Gan2020}, 
albeit with higher computational costs (as discussed below). 
In fact, NLCE have been used to 
study models with discrete disorder, where an exact disorder averaging 
can be performed \cite{Tang2015, Mulanix2019, Park2021}. Moreover, it was 
demonstrated that NLCE 
allow for a more accurate estimation of the critical disorder $W_\ast$ in 
the disordered Heisenberg chain \cite{Devakul2015}. This 
approach was then adapted to study nonequilibrium 
dynamics of inhomogeneous systems \cite{Gan2020}. Building on \cite{Gan2020}, 
we here demonstrate that NLCE can provide insights into the localization 
and delocalization dynamics 
in \text{(quasi-)}1d models, such as ${\cal H}_\text{SL}$ and ${\cal 
H}_\text{FH}$, 
by giving access to the imbalance ${\cal I}(t)$ for $L 
\to \infty$ (see also supplemental material \cite{SuppMat}).  
To this end, consider an infinite system with a random disorder realization, 
and define a unit cell \UCSymbol{scale = 1.2}, e.g., a spin plaquette or two 
neighboring lattice sites, see Fig.~\ref{fig::1}. For a given cluster $c$, 
let ${\cal P}_c(t)$ be the sum \cite{Gan2020}, 
\begin{equation}\label{Eq::NLCE1}
{\cal P}_c(t) = \sum_{{\cal T}(c),\ \UCSymbol{}\subset {\cal T}(c)} 
[{\cal I}_{\UCSymbol{}}(t)]_{{\cal T}(c)}\ ,  
\end{equation}
which runs over all translations 
${\cal T}(c)$ of $c$ such that \UCSymbol{scale = 1.2} is included in 
${\cal T}(c)$, and     
$[{\cal I}_{\UCSymbol{}}(t)]_{{\cal T}(c)}$ denotes the local unit-cell 
imbalance evaluated 
on ${\cal T}(c)$ \cite{Gan2020, SuppMat}. The notion of a cluster here refers 
to a finite part of the full system with OBC. 
Given the (quasi-)1d geometries of ${\cal H}_\text{SL}$ and ${\cal 
H}_\text{FH}$, clusters are 
just ladders or chains of varying size \cite{Mallayya2018, Richter2020, 
NoteLadder} (cf.\ gray rectangles in Fig.\ \ref{fig::1}).
Due to the presence of disorder, $[{\cal I}_{\UCSymbol{}}(t)]_{{\cal T}(c)}$ is 
nonequivalent for different 
translations. The weight of $c$ is then given by an inclusion-exclusion 
principle \cite{Tang2013, Gan2020}, 
\begin{equation}
 {\cal W}_c(t) = {\cal P}_c(t) - \sum_{{\cal T}(c')\subset {\cal T}(c)} 
 {\cal W}_{c'}(t)\ , 
\end{equation}
where the sum runs over all subclusters $c'$ of $c$ (and their translations) 
that include \UCSymbol{scale = 1.2}. The unit cell provides the starting point 
and has no subclusters such that ${\cal W}_{\UCSymbol{scale = 1}}(t) = 
{\cal I}_{\UCSymbol{scale = 1}}(t)$.
The dynamics of the imbalance ${\cal I}(t)$ in the thermodynamic  limit can then 
be approximated as, 
\begin{equation}\label{eq::NLCE3}
\lim_{L\to \infty} {\cal I}(t) \approx \sum_{|c|\leq c_\text{max}} {\cal 
W}_c(t)\ ,  
\end{equation}
including all clusters $c$ up to a cutoff size $|c| = c_\text{max}$ 
that can be handled numerically. While NLCE yield results in the thermodynamic
limit, i.e., there are 
no finite-size effects, one instead has to check the 
convergence of Eq.~\eqref{eq::NLCE3} with respect to the expansion 
order 
$c_\text{max}$, which acts as an effective length scale. 
Typically, a 
larger $c_\text{max}$
leads to convergence on longer time scales \cite{Mallayya2018, Richter2019} 
(or down to lower 
temperatures \cite{Tang2013, Bhattaram2019, 
Schaefer2020}). Reaching large $c_\text{max}$ is 
computationally costly for multiple reasons. First, using full exact 
diagonalization (ED) to evaluate Eqs.\ \eqref{Eq::NLCE1} - \eqref{eq::NLCE3} is 
limited to rather small cluster sizes due to the exponentially growing Hilbert 
space.
Here, we employ an efficient sparse-matrix approach based on 
Chebyshev polynomials \cite{Tal_Ezer_1984, Dobrovitski_2003, Fehske_2009} to 
evaluate $e^{-i{\cal 
H}t}\ket{\psi}$ beyond the 
range of ED \cite{Richter2019, Richter2020}, 
which yields a high accuracy even at long times \cite{SuppMat}. 
Secondly, in the pertinent case of disordered systems, all 
$\sim |c|$ translations of a given 
cluster of size $|c|$ have to be simulated. Due to this computational overhead 
compared to NLCE in translational-invariant models, we (mostly) 
consider expansion orders up to $c_\text{max} = 11$, which 
means that the largest clusters in our simulations 
are ladders of length $L = 11$ (or FH chains with $L = 11$). While 
even larger clusters could in principle be simulated using the 
sparse-matrix approach, we find that this $c_\text{max}$ leads to a 
reasonable tradeoff between the invested computational effort and the 
time scales on which the NLCE remains converged. In addition to 
this main bottleneck of NLCE to reach sufficiently large $c_\text{max}$, 
the costs are further increased by the necessity to perform an average over 
$N_s$ independent disorder samples (here $N_s \approx 10^3$). 
\begin{figure}[tb]
 \centering
 \includegraphics[width=0.95\columnwidth]{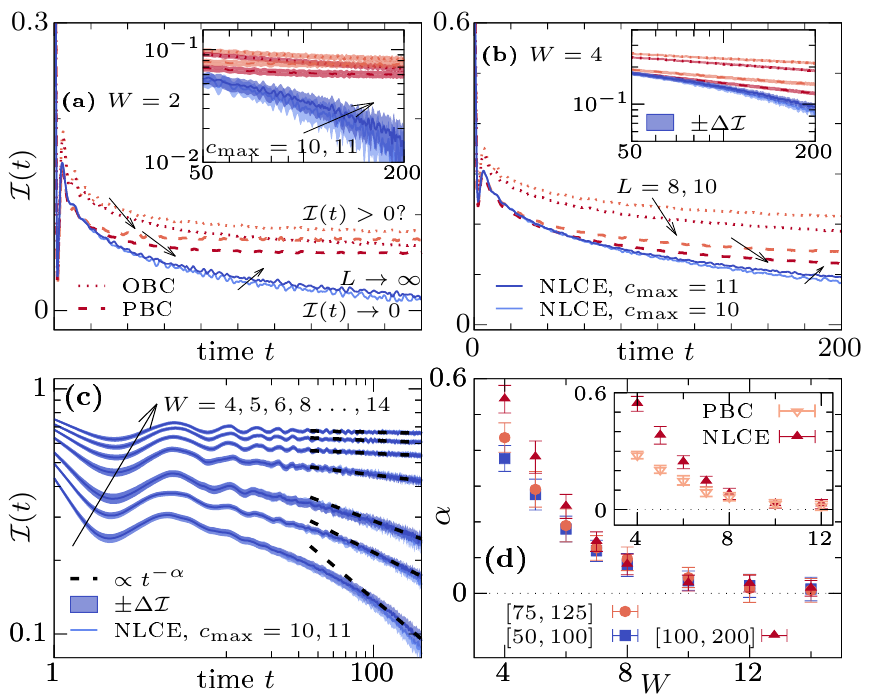}
 \caption{${\cal I}(t)$ in disordered spin 
 ladders~\eqref{eq::Ham_Ladder}. {\bf [(a)(b)]} Data for $W = 2,4$ obtained by 
 NLCE for $c_\text{max} = 10,11$ (solid) are compared to 
simulations of finite systems with $L = 8,10$ with OBC (dotted) and PBC 
(dashed). Direction of increasing $L$ ($c_\text{max}$) is 
indicated by arrows. {\bf Insets:} Same data, 
 but in a double-logarithmic plot. Shaded area indicates standard error $\Delta 
{\cal I}$ of the mean \cite{Note_Error}. 
 {\bf (c)} ${\cal I}(t)$ for different $W$ (arrow). 
At long times, ${\cal I}(t)\propto t^{-\alpha}$. {\bf (d)} 
Exponent $\alpha$, extracted from fitting data in (c) in different time 
windows. {\bf Inset:} $\alpha$ obtained by NLCE ($c_\text{max} = 
10$) and PBC ($L = 10$) for $t \in [100,200]$. 
Data are averaged over $N_s \approx 10^3$ disorder realizations.}
 \label{Fig_Ladder}
\end{figure}

{\it MBL in spin ladders.--}
We now present our numerical results, starting with ${\cal H}_\text{SL}$ and 
the initial state in Eq.~\eqref{Eq::Psi0_Lad}. 
In Figs.\ \ref{Fig_Ladder}~(a) and (b), the imbalance ${\cal I}(t)$ 
is shown for disorder strengths $W = 2$ and $W = 4$. 
Data obtained by NLCE for expansion orders $c_\text{max} = 10,11$ 
(solid curves) are compared to simulations of finite ladders of length $L = 
8,10$ with OBC (dotted) or PBC (dashed). While the $L\to \infty$ dynamics from 
NLCE remain converged up to the longest time $t = 200$ simulated here (see 
\cite{SuppMat} for additional analysis of convergence), we find that ${\cal 
I}(t)$ in the case of $L<\infty$ shows finite-size effects 
already at early times (particularly for OBC). Especially at $W = 2$ [Fig.\ 
\ref{Fig_Ladder}~(a)], ${\cal I}(t)$ obtained by NLCE decays to 
a rather small value, with a slope that indicates that the system 
will delocalize 
at long times. In contrast, in the case of finite systems, ${\cal I}(t)$ decays 
to notably higher values, with the slope of ${\cal 
I}(t)$ being less pronounced. Compared to the $L \to \infty$ NLCE results, 
extrapolating these finite-system data to longer $t$ and larger $L$ is thus more 
intricate and it is less clear whether ${\cal I}(t)$ eventually vanishes. This 
example demonstrates a main result of this Letter. 
In particular, employing NLCE to obtain quantum dynamics for $L \to \infty$ can 
be a powerful means to decide whether a system is 
asymptotically localized or delocalized. Let 
us note that this regime of intermediate disorder is expected to be 
challenging also for other more sophisticated techniques, such as 
matrix-product states, since entanglement presumably still grows rather 
rapidly.

A similar picture also emerges for $W = 4$ 
[Fig.\ \ref{Fig_Ladder}~(b)]. However, as the dynamics are slower 
and finite-size effects are smaller (at least on the time scales shown here), 
the advantage of NLCE compared to direct simulations of finite systems  
becomes less pronounced. Moreover, as emphasized in the insets of Figs.\ 
\ref{Fig_Ladder}~(a) and \ref{Fig_Ladder}~(b), the dynamics of ${\cal I}(t)$ 
obtained by NLCE are more noisy compared to the data for PBC or OBC. This is 
caused by the fact that NLCE relies on the local unit-cell imbalance $[{\cal 
I}_{\UCSymbol{}}(t)]_{{\cal T}(c)}$ \cite{SuppMat}, whereas ${\cal I}(t)$ in 
finite systems is averaged over the full length of the system. 
While the increased noise in the NLCE data may especially 
affect the short-time 
dynamics, we expect it to be less relevant for the qualitative long-time 
behavior of ${\cal I}(t)$.
\begin{figure}[tb]
 \centering
 \includegraphics[width=0.85\columnwidth]{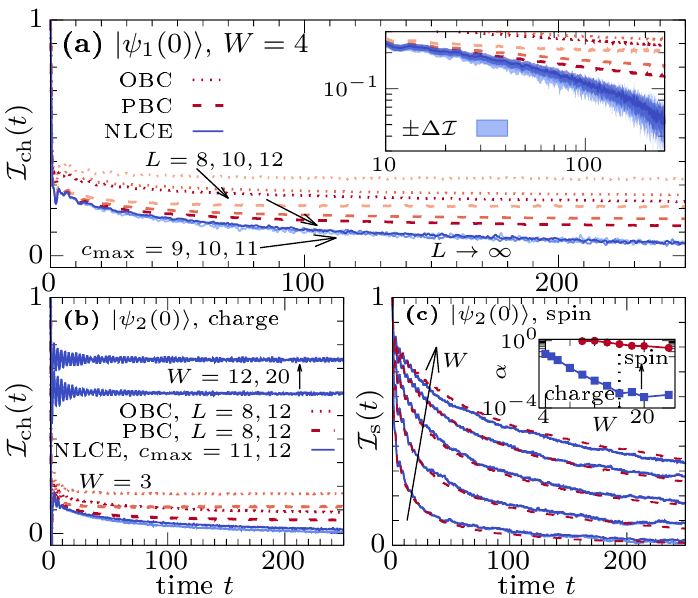}
 \caption{Dynamics in ${\cal H}_\text{FH}$ with $U = 4$ and $V = 0$. {\bf (a)} 
 ${\cal I}_\text{ch}(t)$ for $\ket{\psi_1(0)}$ at $W = 4$, obtained by NLCE for 
$c_\text{max} \leq 11$ (solid) and for finite systems with OBC (dotted) 
and PBC (dashed) and $L = 8,10,12$ (arrow). {\bf Inset:} Same 
data but in a 
double-logarithmic plot. {\bf (b)} ${\cal I}_\text{ch}(t)$ for $\ket{\psi_2(0)}$ 
for 
different $W$. {\bf (c)} ${\cal I}_\text{s}(t)$ for $\ket{\psi_2(0)}$ at $W = 
8,12,\dots,24$ (arrow) using NLCE ($c_\text{max} = 11,12$; 
converged) and systems with OBC ($L = 12$). {\bf Inset:} $\alpha$ obtained by 
fitting ${\cal I}_\text{ch}(t)$ and ${\cal I}_\text{s}(t)$ for  
$t\in[150,250]$.}
 \label{Fig_Hub_1}
\end{figure}

To proceed, Fig.\ \ref{Fig_Ladder}~(c) shows ${\cal I}(t)$ for 
various disorder strengths up to $W = 14$. While the NLCE data for $c_\text{max} 
= 9,10$ remain well converged, we observe that ${\cal I}(t)$ can 
be 
described by a power law \cite{Doggen2018, Sierant2021},
\begin{equation}
 {\cal I}(t) \propto t^{-\alpha}\ , 
\end{equation}
with $\alpha$ depending on $W$. We extract 
$\alpha$ for varying time windows and show the corresponding data in Fig.\ 
\ref{Fig_Ladder}~(d).  
Since one expects $\alpha \to 0$ in the localized phase, 
Fig.\ \ref{Fig_Ladder}~(d) suggests a critical disorder for ${\cal 
H}_\text{SL}$ of $W_\ast \gtrsim 14$, which is notably 
higher 
than for the disordered Heisenberg chain \cite{Luitz2015, Doggen2018, 
Devakul2015}, consistent with ${\cal H}_\text{SL}$ being an intermediate case 
between 1d and 2d \cite{Doggen2020}. As shown in the inset of Fig.\ 
\ref{Fig_Ladder}~(d), extracting $\alpha$ from the dynamics of finite systems 
with PBC and $L = 10$ leads to systematically lower values of $\alpha$ 
(especially for $W \lesssim 8$). Obtaining $\alpha$ from NLCE 
simulations for $L \to \infty$ thus facilitates an accurate estimation of 
$W_\ast$, in line with earlier NLCE studies of eigenstate entanglement 
entropies \cite{Devakul2015}. 

{\it MBL in Fermi-Hubbard chains.--}
We now turn to the dynamics of ${\cal H}_\text{FH}$. We fix 
the interaction to $U = 4$ and, for now, focus on the 
non-tilted model, $V = 0$. 
Figure \ref{Fig_Hub_1}~(a) shows the charge imbalance ${\cal I}_\text{ch}(t)$ 
for 
the initial state $\ket{\psi_1(0)}$ \eqref{Eq::InitState_Hub} at $W = 4$, where 
we again compare the dynamics obtained by NLCE to simulations 
of finite systems with OBC and PBC. Similar to our previous observations 
in the context of ${\cal H}_\text{SL}$, we find that NLCE yield converged 
dynamics on long time scales with a pronounced decay of ${\cal I}_\text{ch}(t)$ 
consistent with delocalization. In contrast, the relaxation of ${\cal 
I}_\text{ch}(t)$ 
for $L < \infty$ is slower and finite-size effects appear at early 
times [cf.\ inset in Fig.\ \ref{Fig_Hub_1}~(a)]. We stress that for the 
highest $c_\text{max} = 11$ considered here, the largest 
clusters are chains with OBC and $L = 11$ \cite{SuppMat}. Nevertheless, 
Fig.\ \ref{Fig_Hub_1}~(a) unveils that combining the contributions of the 
clusters according to Eqs.\ \eqref{Eq::NLCE1} - \eqref{eq::NLCE3} outperforms 
direct simulations of systems with OBC and PBC up to $L = 12$, even though the  
length scales are comparable. NLCE thus proves advantageous also in case of 
moderately disordered FH chains. 
\begin{figure}[tb]
 \centering
 \includegraphics[width = 0.95\columnwidth]{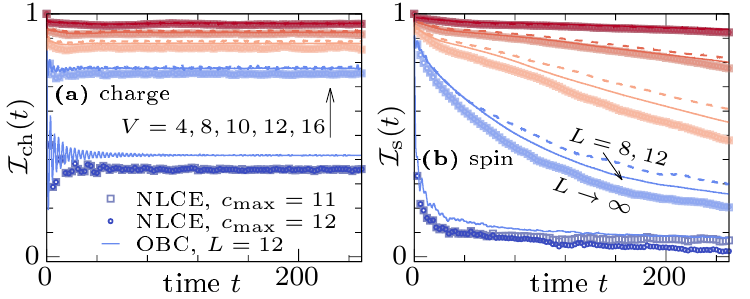}
 \caption{{\bf (a)} ${\cal I}_\text{ch}(t)$ and {\bf (b)} ${\cal I}_\text{s}(t)$ 
for $\ket{\psi_2(0)}$ at fixed $W = 4$, $U = 4$, and different 
lattice tilts $V\leq 16$. Data for chains with OBC and $L = 8, 12$ 
(dashed, solid) are compared to $L \to \infty$ NLCE dynamics with $c_\text{max} 
= 11,12$ (squares, circles). NLCE remains well converged except for ${\cal 
I}_\text{s}(t)$ and $V = 4$. 
   }
 \label{Fig_Tilt}
\end{figure}

Next, we consider the initial state $\ket{\psi_2(0)}$ in 
Eq.\ \eqref{Eq::Init::Hub2} with ${\cal I}_\text{ch}(t)$ shown in 
Fig.\ \ref{Fig_Hub_1}~(b) for exemplary values of $W$. While we observe 
delocalized dynamics for $W = 3$, most pronounced in case of the 
$L \to \infty$ NLCE data, ${\cal I}_\text{ch}(t)>0$ approaches approximately 
time-independent plateaus for $W = 12,20$, suggesting charge 
localization at sufficiently strong disorder, cf.\ inset in Fig.\ 
\ref{Fig_Hub_1}~(c). While  
nonergodic charge dynamics has been observed before \cite{Prelovsek2016, 
Mondaini2015}, spin was found to be delocalized and relax subdiffusively instead 
\cite{Kozarzewski2018}. Here, we explore the fate of spin dynamics 
for $L \to \infty$ at strong disorder. Specifically, the spin imbalance 
${\cal I}_\text{s}(t)$ is shown in Fig.\ \ref{Fig_Hub_1}~(c) for $W \leq 24$. 
Using 
NLCE up to $c_\text{max} = 12$ \cite{Note_NLCE_Psi2}, we find no signatures of 
localization and ${\cal I}_\text{s}(t)$ decays as $\propto t^{-\alpha}$, 
$\alpha > 0$ 
[inset in Fig.\ \ref{Fig_Hub_1}~(c)]. The fact that the $L \to \infty$ dynamics 
obtained by 
NLCE and for systems with $L = 12$ agree very 
well with each other demonstrates that finite-size effects are negligible. Spin 
dynamics of ${\cal H}_\text{FH}$ thus behave 
delocalized even at extremely large $W$, where charge is frozen.    

{\it Dynamics in tilted lattice.--}
We now consider ${\cal H}_\text{FH}$ with $V > 
0$. While such tilts may lead to Stark localization 
\cite{Schulz2019, vanNieuwenburg2019, Guo2021}, nonergodic dynamics in strongly 
tilted 
lattices has also been attributed to Hilbert-space 
fragmentation \cite{Scherg2021, Khemani2020, Sala2020, Doggen2021}. Fixing the 
disorder to 
$W = 4$ (for which ${\cal H}_\text{FH}$ is delocalized 
at $V = 0$, cf.\ Fig.\ \ref{Fig_Hub_1}), Fig.\ \ref{Fig_Tilt} shows ${\cal 
I}_\text{ch}(t)$ and ${\cal I}_\text{s}(t)$ resulting from quenches with 
$\ket{\psi_2(0)}$ and 
different $V>0$, obtained by NLCE for $L \to \infty$ and 
direct simulations of 
finite chains with OBC. While $W>0$ is convenient 
to suppress the strong oscillations of ${\cal I}(t)$ \cite{SuppMat}, 
the combination of disorder and lattice tilt may reinforce localization 
\cite{vanNieuwenburg2019}. 
Remarkably, we observe in Fig.\ \ref{Fig_Tilt} that not only ${\cal 
I}_\text{ch}(t)$ ceases to 
decay with increasing $V$, but ${\cal I}_\text{s}(t)$    
also slows down drastically with $V$, especially compared to the case 
of bare disorder and no tilt [cf.\ Fig.\ \ref{Fig_Hub_1}~(c)]. In particular, 
for the largest $V = 16$ considered here, ${\cal I}_\text{s}(t)$ does not 
substantially decay for $t < 250$, suggesting the possibility to induce 
nonergodic spin dynamics on experimentally relevant time scales in tilted 
lattices. This is another 
key result. We stress that this mechanism of causing slow 
spin 
dynamics is distinct from other 
examples where localization was achieved by lifting the SU$(2)$ symmetry of 
${\cal H}_\text{FH}$  
\cite{Sroda2019}. While 
${\cal I}_\text{s}(t)$ appears to be strongly initial-state dependent 
(see \cite{SuppMat}), we here 
leave it to future work to explore 
the effect of $V > 0$ in more detail.

{\it Conclusion.--} 
To summarize, we have employed NLCE 
to study quantum quenches in disordered spin ladders and 
Fermi-Hubbard 
chains and obtained converged results for the imbalance ${\cal I}(t)$ on 
comparatively long time scales. 
We have put particular emphasis on intermediate disorder values $W < W_\ast$, 
where we demonstrated that NLCE outperform direct simulations of finite 
systems 
with OBC or PBC. 
Furthermore, in contrast to bare disorder, our 
analysis predicts that an additional tilted potential leads to a notable
slowdown of spin dynamics for certain initial states in FH chains, 
which should be accessible 
experimentally \cite{Scherg2021}.
Even though NLCE yield results for $L \to \infty$, allowing better 
estimates for $W_\ast$ \cite{Devakul2015}, we stress 
that, similar to other methods, an unambiguous detection of MBL is beyond its  
capabilities (and was not our goal). Since simulation times are limited, $t < 
\infty$, extracted values for $W_\ast$ should be understood as lower bounds 
for 
the putative MBL transition. 

Given the apparent advantage of NLCE at intermediate disorder, 
a natural direction of research is to explore the emergence of 
subdiffusion on the ergodic side of the MBL transition \cite{Luitz2017, 
Weiner2019, Varma2017, Richter2018}. In this context, NLCE have been shown to be 
a 
powerful means to study transport properties of 1d systems in the thermodynamic 
limit \cite{Richter2019}.     
Another interesting avenue is to consider MBL in higher 
dimensions. While NLCE have proven competitive with other state-of-the-art 
methods to simulate 
quantum dynamics in 2d \cite{Richter2020_2}, reaching 
high expansion orders 
for 2d lattices is computationally demanding such that convergence times are  
still limited \cite{Gan2020}.    

{\it Acknowledgements.--} 
This work was funded by the 
European Research 
Council 
(ERC) under the European Union's Horizon 2020 research and innovation programme
(Grant agreement No.\ 853368).

\clearpage
\newpage
\onecolumngrid

% %%%%%%%%%%%%%%%%%%%%%%%%%
% % Supplemental Material %
% %%%%%%%%%%%%%%%%%%%%%%%%%
 
\setcounter{figure}{0}
\setcounter{equation}{0}
\renewcommand*{\citenumfont}[1]{S#1}
\renewcommand*{\bibnumfmt}[1]{[S#1]}
\renewcommand{\thefigure}{S\arabic{figure}}
\renewcommand{\theequation}{S\arabic{equation}}

\section*{Supplemental material}

\section{Numerical linked cluster expansion for disordered 1d systems}

\begin{figure}[h]
 \centering
 \includegraphics[width=0.9\textwidth]{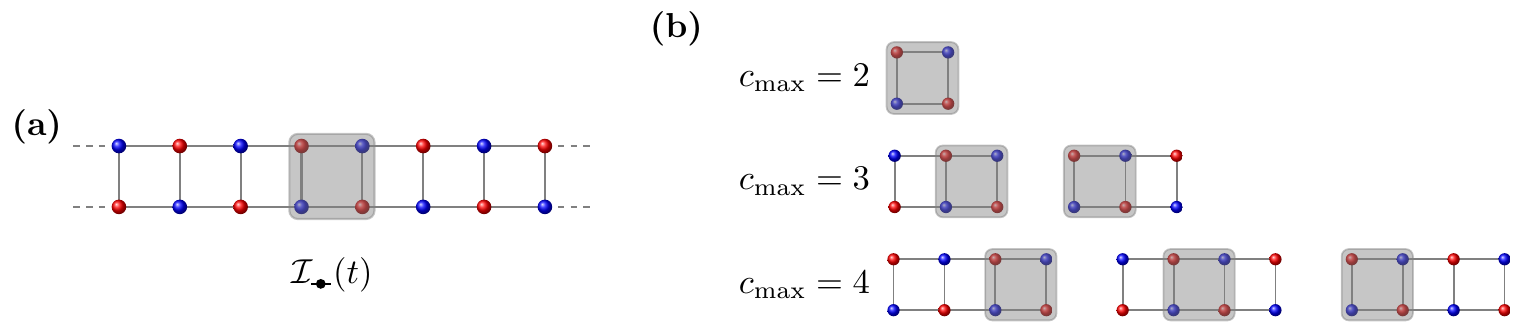}
  \caption{{\bf (a)} NLCE provide a means to study the dynamics of the 
  imbalance ${\cal I}(t)$ in the thermodynamic limit $L \to \infty$. To this 
end, consider an infinite ladder with inhomogeneous initial state (here N\'eel 
state as indicated by the red and blue lattice sites) and a fixed realization 
of disorder (not shown here). Next, identify a suitable unit cell to calculate 
the local unit-cell imbalance ${\cal 
I}_{-\hspace{-0.17cm}\bullet}(t)$ (here a plaquette of 
length $L = 2$ with $2L = 4$ lattice sites in total, as indicated by the shaded 
area). 
   {\bf (b)} ${\cal I}(t)$ in the thermodynamic limit is then obtained 
   by suitably combining the contributions of clusters $c$ [and their 
translations ${\cal T}(c)$], which contain the unit cell (see \cite{Gan2020S} 
and main text for details). For the quasi one-dimensional ladder, we consider 
only clusters that are ladders as well \cite{Richter2020_2S}. All clusters that 
have to be evaluated up to expansion order $c_\text{max} = 4$ are shown, where 
$c_\text{max}$ refers to the length of the largest cluster. Including larger 
and 
larger expansion orders leads to converged results of ${\cal I}(t)$ on longer 
time scales.}
 \label{Fig::NLCE_1d}
\end{figure}

In addition to our explanations in the main text, let us provide further 
details on how to set up NLCE for disordered systems with inhomogeneous initial 
states, see also \cite{Gan2020S}. The starting point is provided by a finite 
section of the infinite system, with a fixed disorder realization and initial 
state. As shown in Fig.\ \ref{Fig::NLCE_1d}~(a), one then selects a unit cell 
\UCSymbol{scale = 1.2}, which in our case is a single spin plaquette (in case 
of 
the spin ladder) or two neighboring lattice sites (in case of the Hubbard 
chain). While the imbalance ${\cal I}(t)$ is defined as a sum over the full 
system, the NLCE formalism relies on the calculation of the local unit-cell 
imbalance ${\cal I}_{\UCSymbol{}}(t)$, i.e., ${\cal I}(t)$ restricted to the 
chosen unit cell. As described in the main text, NLCE then consists of 
simulating finite clusters of increasing size and combining their contributions 
suitably. In Fig.\ \ref{Fig::NLCE_1d}~(b), we plot all clusters that have to be 
evaluated up to expansion order $c_\text{max} = 4$, which means 
that the largest cluster in the expansion is of size $4 \times 2$. In 
particular, for each cluster $c$ with a given size, all its translations ${\cal 
T}(c)$ shifted around the unit cell have to be evaluated. Note that for 
simulating the contributions of these clusters, the disorder realization and 
the 
respective alignment of the initial state has to remain fixed.     
Since the translations therefore contain different parts of the static disorder 
configuration, ${\cal I}_{\UCSymbol{}}(t)$ can vary for different translations. 
While NLCE by construction yields results in the thermodynamic $L\to\infty$, it 
is crucial to check the convergence of the series. Increasing $c_\text{max}$, 
i.e., including clusters with longer length scales, will typically increase the 
time scales on which the dynamics of ${\cal I}(t)$ remains converged.  

As becomes apparent from Fig.\ \ref{Fig::NLCE_1d}~(b), the computational 
costs of employing NLCE up to some expansion order $c_\text{max}$ are notably 
higher than a direct simulation of a finite system with PBC or OBC of length $L 
= c_\text{max}$. In particular, while the latter requires 
the simulation of merely a single finite system (multiplied by the number of 
desired disorder samples), the former requires the simulation of multiple 
finite 
clusters within each expansion order, which is polynomially (roughly by a 
factor 
of $L$) more costly. As a consequence, we here restrict ourselves to expansion 
orders $c_\text{max} \lesssim 11$ [or $c_\text{max} \lesssim 12$ in case of 
Fermi-Hubbard chains with quarter-filling initial state $\ket{\psi_2(0)}$ in 
Eq.\ (5)]. While it is certainly possible to evaluate ${\cal 
I}(t)$ using sparse-matrix techniques on even larger clusters, NLCE simulations 
for this value of $c_\text{max}$ are already quite demanding and, in practice, 
yield converged results on sufficiently long time scales.  

Eventually, as also mentioned in the main text, NLCE typically require 
disorder averaging over a larger number of samples to yield the same noise 
level 
as direct simulations of finite systems with PBC or OBC. This is caused by the 
fact that NLCE relies on the local evaluation of ${\cal I}_{\UCSymbol{}}(t)$ on 
the unit-cell, which is significantly smaller than the full system. In 
particular, while ${\cal I}_{\UCSymbol{}}(t)$ is in our case defined on $2$ or 
$4$ lattice sites (and is therefore most sensitive to the random 
fields/potentials on these sites), the imbalance ${\cal I}(t)$ in the case of 
finite $L$ is given by a sum over the full system, which leads to an effective 
disorder averaging over $\propto L$ local random fields/potentials. 

\section{Pure-state propagation}

\subsection{Chebyshev polynomial expansion}

In order to access system (and cluster) sizes beyond the range of 
exact diagonalization (ED), we here subdivide the evolution up to 
time $t$ into a product of discrete 
time steps,  
\begin{equation}
\ket{\psi(t)} = e^{-i{\cal H}t}\ket{\psi(0)} =  \left(e^{-i{\cal H}\delta 
t}\right)^Q \ket{\psi(0)}\ ,  
\end{equation}
where $\delta t = t/Q$. We approximate the 
action of the exponential $\exp(-i{\cal H}\delta t)$ by a Chebyshev-polynomial 
expansion \cite{Tal_Ezer_1984S, Dobrovitski_2003S, Fehske_2009S}. 
Since the Chebyshev polynomials are defined on the interval $[-1,1]$, the 
spectrum of the original Hamiltonian ${\cal H}$ has to be rescaled 
\cite{Fehske_2009S},  
\begin{equation}
 \widetilde{{\cal H}} = \frac{{\cal H} - b}{a}\ , 
\end{equation}
where $a$ and $b$ are suitably chosen parameters. In practice, we use the fact 
that the (absolute of the) extremal eigenvalue of ${\cal H}$ can be bounded 
from above \cite{Dobrovitski_2003S}. For instance, in case of the
disordered spin ladder ${\cal H}_\text{SL}$ we have,   
\begin{equation}
\max(|E_\text{min}|,|E_\text{max}|) \leq \left(3N_{\langle 
\ell,m\rangle}/4 + 2  W L/2 \right) = {\cal E}\ ,
\end{equation}
where $E_\text{max}$ ($E_\text{min}$) is the largest (smallest) eigenvalue of 
${\cal H}_\text{SL}$, $N_{\langle 
\ell,m\rangle} = 2(L-1) + L$ denotes the number of nearest-neighbor 
bonds (in the case of OBC), and the disorder $W$ couples to $2L$ operators 
$S_\ell^z$ with maximal eigenvalue $1/2$ each. 
Similar bounds for $E_\text{max}$ ($E_\text{min}$) can be obtained for 
disordered Hubbard chains ${\cal H}_\text{FH}$ as well. 
By choosing 
$a \geq {\cal E}$, it is guaranteed that the spectrum of ${\widetilde{\cal 
H}}$ lies within $[-1,1]$. As a consequence, we can set $b = 0$. Note 
that while this choice of $a$ and $b$ is not necessarily 
optimal, it proves to be sufficient \cite{Dobrovitski_2003S}.

Within the Chebyshev-polynomial formalism, the time evolution of a state 
$\ket{\psi(t)}$ can then be approximated as an 
expansion up to order $M$ \cite{Fehske_2009S}, 
\begin{equation}\label{Eq::Cheb1}
 \ket{\psi(t+\delta t)} \approx c_0\ket{v_0} + \sum_{k=1}^M 2c_k\ket{v_k}\ , 
\end{equation}
where the expansion coefficients $c_0, c_1, \dots, c_M$,  
are given by
\begin{equation}\label{Eq::ChebCoeff}
 c_k = (-i)^k {\cal J}_k(a\delta t)\ , 
\end{equation}
with ${\cal J}_k(a\delta t)$ being the $k$-th order Bessel function of the 
first 
kind evaluated at $a\delta t$. 
[Note that Eqs.\ \eqref{Eq::Cheb1} and \eqref{Eq::ChebCoeff} 
assume $b = 0$.] Moreover, the vectors 
$\ket{v_k}$ 
are recursively generated according to 
\begin{equation}\label{Eq::Cheb2}
 \ket{v_{k+1}} = 2\widetilde{{\cal H}}\ket{v_k} - \ket{v_{k-1}}\ ,\ \quad k 
\geq 1\ , 
\end{equation}
with $\ket{v_1} = \widetilde{{\cal H}}\ket{v_0}$ and $\ket{v_0} = 
\ket{\psi(t)}$. Given a time step $\delta t$ (and the parameter $a$), 
the expansion order $M$ has to be chosen large enough to ensure negligible 
numerical errors. 

As becomes apparent from Eqs.\ \eqref{Eq::Cheb1} and \eqref{Eq::Cheb2}, the 
time evolution of the pure state $\ket{\psi(t)}$ requires the evaluation of 
matrix-vector products. Since ${\widetilde{\cal H}}$ is a sparse matrix, these 
matrix-vector multiplications can be implemented comparatively time and 
memory efficient. As a consequence, it is possible to treat system 
(or cluster) sizes 
that are larger compared to ED.

\subsection{Accuracy of the time evolution}

\begin{figure}[tb]
 \centering
 \includegraphics[width=0.9\textwidth]{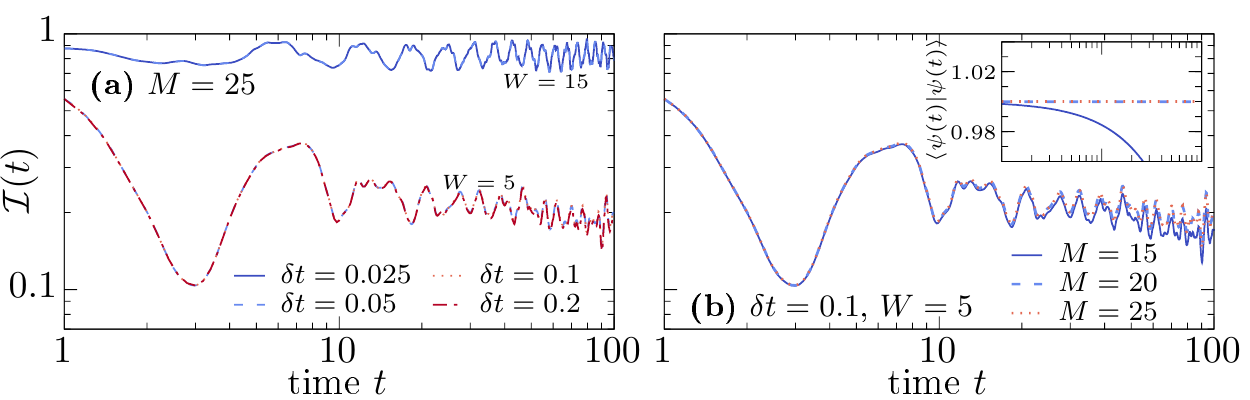}
 \caption{{\bf (a)} Imbalace ${\cal I}(t)$ in spin ladders with
 OBC of length $L = 10$ and disorder $W = 5,15$. Data for different time steps 
$\delta t$ is compared at fixed Chebyshev expansion order $M = 25$. {\bf (b)} 
${\cal I}(t)$ for $W = 5$ using $\delta t = 0.1$ and different choices of $M$. 
The inset shows the corresponding time dependence of the norm of 
$\ket{\psi(t)}$.}
 \label{Fig::Time}
\end{figure}

In Fig.\ \ref{Fig::Time}~(a), the dynamics of the 
imbalance ${\cal I}(t)$ in a spin ladder of length $L = 10$ with 
open boundary conditions is shown for $W = 5$ and $W = 15$, using a 
single realization of disorder in both cases. We compare data for different 
time 
steps $\delta t$ at fixed Chebyshev expansion order $M = 25$. For all values of 
$\delta t$ considered here, we find that the resulting dynamics of ${\cal 
I}(t)$ 
is practically independent of $\delta t$ (note that for $W = 15$, we only 
consider the smaller $\delta t = 0.025,0.05$.). Focusing on $W = 5$, Fig.\ 
\ref{Fig::Time}~(b) furthermore shows results for a fixed time step $\delta t = 
0.1$ and varying values of $M$. While the data for $M = 20$ and $M = 25$ agree 
convincingly, deviations become apparent for the smaller $M = 15$. 
In the inset of Fig.\ \ref{Fig::Time}~(b), we additionally depict 
the norm $||\ket{\psi(t)}||^2 = \braket{\psi(t)|\psi(t)}$. For $M = 20,25$, we 
find that $||\ket{\psi(t)}||$ remains essentially constant, $||\ket{\psi(t)}|| 
= 
1$, for the entire time window considered here, whereas the conservation of 
$||\ket{\psi(t)}||$ is violated for $M = 15$.     
In view of the data shown in Fig.\ \ref{Fig::Time}, the 
numerical data presented in this paper are obtained using a time step $\delta t 
= 0.1$ for low to intermediate disorder $W\leq 6$, while we choose $\delta t = 
0.05$ for larger $W$ (and sometimes even smaller $\delta t$ for large $W$ and 
$V$). We fix $M = 25$. Using 
these parameters, we obtain 
accurate data on long time scales, which is crucial such that the NLCE remains 
well controlled when combining the contributions of multiple clusters. 

\section{Convergence of NLCE}

\begin{figure}[tb]
 \centering
 \includegraphics[width=0.9\textwidth]{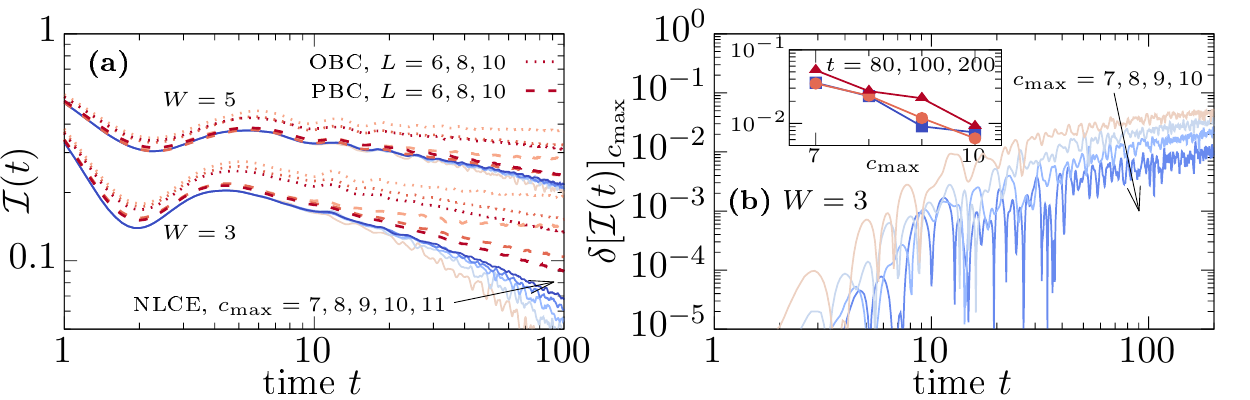}
 \caption{{\bf (a)} Decay of the imbalance ${\cal I}(t)$ for ${\cal 
H}_\text{SL}$ 
 with $W = 3,5$, obtained by NLCE for expansion orders $c_\text{max} = 7-11$. 
As a comparison, we show data obtained in finite systems with PBC and OBC and 
$L 
= 6,8,10$. {\bf (b)} Difference $\delta[{\cal I}(t)]_{c_\text{max}}$ between 
the dynamics of ${\cal I}(t)$ obtained for different expansion orders 
$c_\text{max}$ [cf.\ Eq.\ \eqref{Eq::Diff}] at $W = 3$. {\bf Inset:} 
$\delta[{\cal I}(t)]_{c_\text{max}}$ at fixed times $t = 80,100,200$ versus 
$c_\text{max}$.}
 \label{Fig::Convergence}
\end{figure}

As described in the main text, increasing the maximum cluster size 
$c_\text{max}$ will typically increase the time scales on which NLCE yield 
converged results. Focusing on ${\cal H}_\text{SL}$, Fig.\ 
\ref{Fig::Convergence}~(a) 
shows ${\cal I}(t)$ for two different disorder strengths $W = 3$ and $ W = 5$, 
obtained by NLCE for $c_\text{max} = 7 - 11$. Moreover, we show data for finite 
systems with PBC or OBC of length $L = 6,8,10$. Comparing the two values of 
$W$, 
we find that NLCE remain converged on longer time scales if disorder is 
stronger. In particular, while the curves for different $c_\text{max}$ agree 
convincingly with each other for $W = 5$ up to the longest times $t \leq 200$ 
shown here, a breakdown of convergence can be clearly seen in the case of $W = 
3$ and smaller $c_\text{max}$. This observation can be understood by the fact 
that the dynamics become
more and more localized for stronger disorder, i.e., the relevant length scales 
become shorter, such that clusters of smaller size are able to capture the 
dynamics in the thermodynamic limit. Importantly, while the NLCE data for $W = 
5$ are well converged for all $c_\text{max}$ shown here, the corresponding data 
of ${\cal I}(t)$ for finite systems with OBC or PBC in Fig.\ 
\ref{Fig::Convergence}~(a) still show distinct finite-size effects.   
This is in line with our findings from the main text, i.e., for a given 
$c_\text{max}$ (corresponding to finite systems of length $L = c_\text{max}$), 
NLCE yield a better convergence than direct simulations of finite systems.   

To gain more insights into the convergence properties of NLCE, let us define, 
\begin{equation}\label{Eq::Diff}
 \delta[{\cal I}(t)]_{c_\text{max}} = |[{\cal I}(t)]_{c_\text{max}=11} - 
 [{\cal I}(t)]_{c_\text{max}} |\ , 
\end{equation}
which is the difference between ${\cal I}(t)$ obtained by NLCE for some 
expansion order $c_\text{max}<11$ and ${\cal I}(t)$ obtained for the largest 
$c_\text{max} =11$ that is available to us. Focusing on $W = 3$, Fig.\ 
\ref{Fig::Convergence}~(b) shows $\delta[{\cal I}(t)]_{c_\text{max}}$ for 
$c_\text{max} = 7-10$, as obtained from the data of ${\cal I}(t)$ in Fig.\ 
\ref{Fig::Convergence}~(a). While $\delta[{\cal I}(t)]_{c_\text{max}}$ 
essentially vanishes at short times (i.e., the NLCE is well converged), we find 
that $\delta[{\cal I}(t)]_{c_\text{max}}$ grows with increasing time. While 
this 
indicates that the convergence of NLCE becomes worse at longer times, we find 
in 
Fig.\ \ref{Fig::Convergence}~(b) that $\delta[{\cal I}(t)]_{c_\text{max}}$ 
systematically decreases with increasing $c_\text{max}$. In particular, as 
emphasized in the inset of Fig.\ \ref{Fig::Convergence}~(b), $\delta[{\cal 
I}(t)]_{c_\text{max}}$ at fixed times $t = 80,100,200$ decreases 
approximately exponentially with $c_\text{max}$. This demonstrates that the 
convergence of NLCE can be improved in a controlled way by including higher and 
higher expansion orders.  

\section{Decay of imbalance in strongly disordered Hubbard chains}

\begin{figure}[h]
 \centering
 \includegraphics[width=0.9\textwidth]{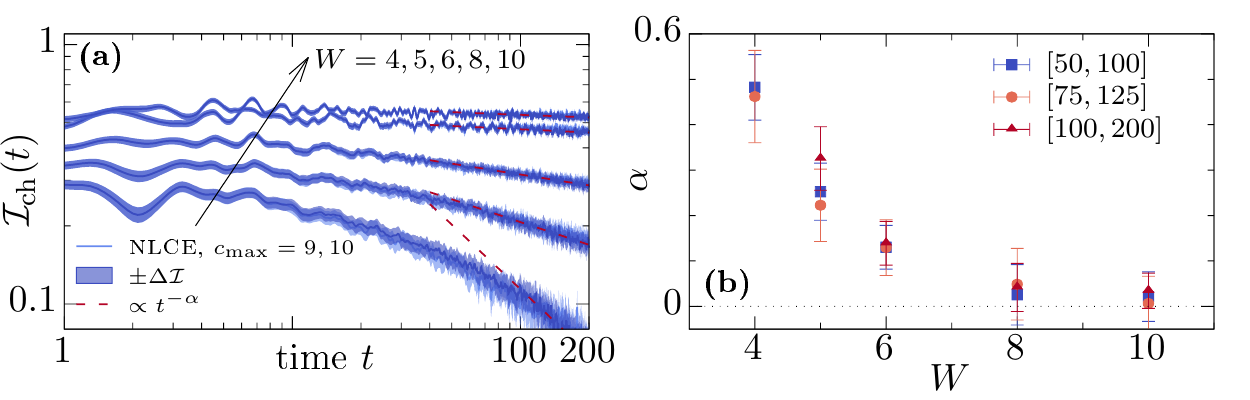}
 \caption{{\bf (a)} Charge imbalance ${\cal I}_\text{ch}(t)$ in Fermi-Hubbard 
 chains with initial state $\ket{\psi_1(0)}$ [Eq.\ (4)] 
and various disorder values $W$ (arrow), obtained by NLCE for expansion orders 
$c_\text{max} = 9,10$. Shaded area indicates the standard error of the mean. 
{\bf (b)} Power-law exponent $\alpha$ extracted from fits ${\cal 
I}_\text{ch}(t) 
\propto 
t^{-\alpha}$ [cf.\ dashed curves in panel (a)] in different time windows. We 
have $U = 4$ in all cases.}
 \label{Fig2_Hub}
\end{figure}

Let us present additional data for the decay of the charge imbalance 
${\cal I}_\text{ch}(t)$ in Fermi-Hubbard chains 
${\cal H}_\text{FH}$ at strong disorder. Focusing on the initial state 
$\ket{\psi_1(0)}$, Fig.\ \ref{Fig2_Hub} shows ${\cal I}_\text{ch}(t)$ obtained 
by NLCE 
for $c_\text{max} = 9,10$ for varying disorder strengths $W \leq 10$. Similar 
to 
the case of the spin ladder considered in Fig.\ 2 in the main 
text, we find that ${\cal I}_\text{ch}(t) \propto t^{-\alpha}$ can be fitted by 
a power 
law at long times. In Fig.\ \ref{Fig2_Hub}~(b), we plot $\alpha$ versus $W$. 
While $\alpha$ appears to approach zero for sufficiently strong $W$, the data 
in 
Fig.\ \ref{Fig2_Hub}~(b) suggests that the critical disorder $W_\ast$ (for the 
half-filling sector probed by $\ket{\psi_1(0)}$) is probably even larger than 
the strongest value of $W$ considered here, $W_\ast \gtrsim 10$. Let us stress 
that 
this estimate is based on finite-time data $t < 200$, such that we cannot make 
statements about the fate of charge localization in the limit $t\to \infty$ and 
the potential impact of the thermalizing spin dynamics [cf.\ Fig.\ 
3~(c) in main text]. Moreover, we note that the extraction of 
$W_\ast$ can depend on the chosen initial state and its properties such as the 
density of doublons and singlons \cite{Protopopov2019S}.  

\section{Additional data for decay of the spin imbalance in 
tilted Fermi-Hubbard chains}

Let us present additional data for the dynamics of ${\cal I}_\text{s}(t)$ in 
Fermi-Hubbard chains with $V > 0$. In contrast to Fig.\ 2 of the 
main text, where we considered the dynamics resulting from $\ket{\psi_2(0)}$ 
[cf.\ Eq.\ (5)], we now study two different initial states,
\begin{align}
 \ket{\psi_3(0)} &= \prod_{\ell} c_{2\ell,\uparrow}^\dagger 
c_{2\ell+1,\downarrow}^\dagger\ket{0} =    \left\vert \begin{array}{cccccc}
\cdots & \uparrow\ & \downarrow\ & \uparrow\ & \downarrow\  & 
\cdots 
\end{array}\right\rangle\ ,  \label{Eq::Psi3} \\ 
\ket{\psi_4(0)} &= \prod_{\ell} c_{4\ell,\uparrow}^\dagger 
c_{4\ell+1,\uparrow}^\dagger 
c_{4\ell+2,\downarrow}^\dagger c_{4\ell+3,\downarrow}^\dagger\ket{0} =    
\left\vert \begin{array}{cccccccccc}
\cdots & \downarrow\ & \downarrow\ & \uparrow\ & \uparrow\ & \downarrow\ & 
\downarrow\ & \uparrow\ & \uparrow\ & 
\cdots 
\end{array}\right\rangle\ \label{Eq::Psi4}. 
\end{align}
In contrast to $\ket{\psi_2(0)}$, which has quarter-filling, $\ket{\psi_3(0)}$ 
and $\ket{\psi_4(0)}$ 
probe the dynamics in the half-filling sector. Taking the example of 
$\ket{\psi_3(0)}$ it is also insightful to write the spin imbalance ${\cal  
I}_\text{s}(t)$  as, 
\begin{equation}
 {\cal I}_\text{s}(t) \propto \sum_{\ell} (-1)^{\ell} \langle 
(n_{\ell,\uparrow} - n_{\ell,\downarrow})(t) \rangle = \sum_{\ell = 1}^L 
\langle (n_{\ell,\uparrow} - n_{\ell,\downarrow})(t)\rangle 
\langle (n_{\ell,\uparrow} - n_{\ell,\downarrow})(0)\rangle = \sum_{\ell = 1}^L 
\langle m_\ell(t)\rangle \langle m_\ell(0)\rangle\ , 
\end{equation}
where we have rewritten ${\cal I}_\text{s}(t)$ as a correlation function and 
introduced 
the local magnetization $m_\ell = n_{\ell,\uparrow} - n_{\ell,\downarrow}$. We 
note that the imbalances ${\cal I}_\text{ch}(t)$ and ${\cal I}_\text{s}(t)$ 
considered in the main text for $\ket{\psi_1(0)}$ and $\ket{\psi_2(0)}$ can 
be written in the form of similar correlation functions as well.  

\begin{figure}[tb]
 \centering
 \includegraphics[width = 0.85\textwidth]{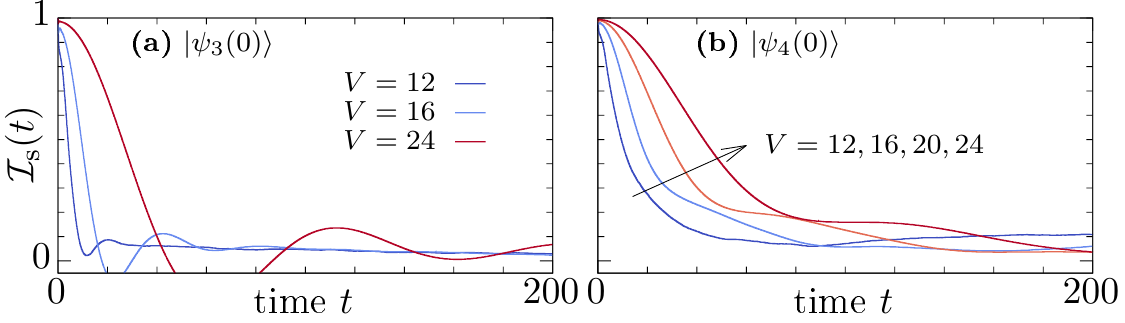}
 \caption{Spin imbalance in ${\cal H}_\text{FH}$ for different tilt strengths 
$V$, resulting from initial 
state {\bf (a)} $\ket{\psi_3(0)}$ [Eq.\ \eqref{Eq::Psi3}] and {\bf (b)} 
$\ket{\psi_4(0)}$ [Eq.\ \eqref{Eq::Psi4}]. Data are obtained for finite systems 
with $L = 10$ and OBC.   We have $U = 4$ 
and $W = 4$ in all cases. }
 \label{Fig_S_Psi3}
\end{figure}

In Figs.\ \ref{Fig_S_Psi3}~(a) and (b), ${\cal I}_\text{s}(t)$ resulting from 
the initial 
states $\ket{\psi_3(0)}$ and $\ket{\psi_4(0)}$ is shown for fixed $U = W = 4$ 
and varying lattice 
tilts $V$. While ${\cal I}_\text{s}(t)$ is found to decay towards zero in both 
cases, indicating delocalization of spin degrees of freedom even for strong 
lattice tilts $V = 24$, we find that ${\cal I}_\text{s}(t)$ does depend notably 
on the initial state. In particular, in the case of $\ket{\psi_4(0)}$ [Fig.\ 
\ref{Fig_S_Psi3}~(b)], the decay of ${\cal I}_\text{s}(t)$ is found to be less 
abrupt and can be further slowed down by increasing $V$. Both for 
$\ket{\psi_3(0)}$ and $\ket{\psi_4(0)}$, however, the dynamics of ${\cal 
I}_\text{s}(t)$ are distinctly faster compared to the example of 
$\ket{\psi_2(0)}$ considered in Fig.\ 4 in the main text. 

\section{Dynamics in the tilted Fermi-Hubbard chain without 
disorder}

\begin{figure}[tb]
 \centering
 \includegraphics[width=0.85\textwidth]{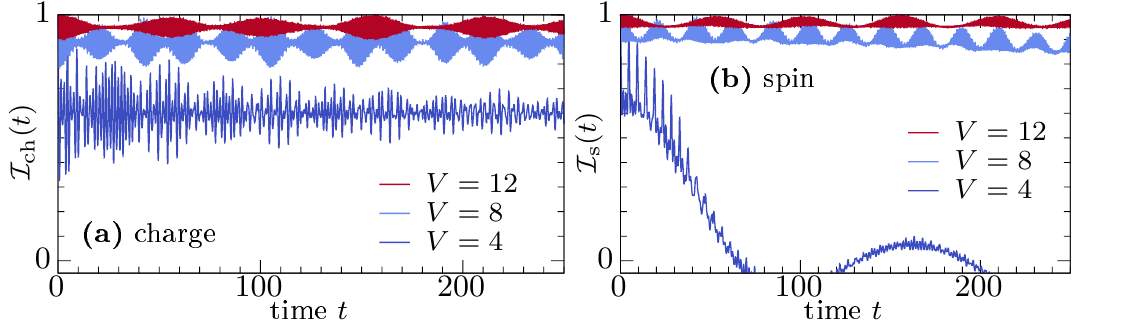}
 \caption{{\bf (a)} Charge imbalance ${\cal I}_\text{ch}(t)$ in the 
Fermi-Hubbard model ${\cal H}_\text{FH}$ without disorder ($W = 0$), resulting 
from quenches with the initial 
state $\ket{\psi_2}$ [see Eq.\ (5) in main text] and 
different lattice tilts $V = 4,8,12$. Data are shown for chains 
with 
$L = 12$ and open boundary conditions. {\bf (b)} Analogous data, but now for 
the spin imbalance ${\cal I}_\text{s}(t)$.}
 \label{Fig_Tilt_W0}
\end{figure}

While we have considered the dynamics of ${\cal H}_\text{FH}$ 
in the main text for either $W > 0$ at $V = 0$ (Fig.\ $3$) or for $V > 0$ at $W 
= 4$ (Fig. $4$), let us here present additional data for the case of having 
just a tilted lattice without additional disorder, i.e., $V > 0$ and $W = 0$. 
In Figs.\ \ref{Fig_Tilt_W0}~(a) and (b), we show the charge and spin imbalances 
${\cal 
I}_\text{ch}(t)$ and ${\cal I}_\text{s}(t)$ resulting from quenches with the 
initial state $\ket{\psi_2}$ [Eq.\ $(5)$ in main text] and different tilt 
values $V 
> 0$. The data are obtained for finite systems with $L = 12$ and open boundary 
conditions. To begin with, for $V = 4$, we find that while ${\cal 
I}_\text{ch}(t\to\infty)> 0$ 
saturates to a finite, 
approximately constant, long-time value, the spin imbalance ${\cal 
I}_\text{s}(t)$ clearly decays towards zero. For larger $V = 8$ and $V = 12$, 
in contrast, 
we find that spin dynamics clearly slow down as well. In particular, for $V 
=12$, we are unable to observe any notable decay of ${\cal I}_\text{s}(t)$ on 
the time scale $t \leq 250$ shown here. While the data in Fig.\ 
\ref{Fig_Tilt_W0} is qualitatively similar to the data shown in Fig.\ 4 in the 
main text, we also note a number of differences. In particular, compared to 
the results of 
${\cal I}_\text{s}(t)$ in Fig.\ 4~(b), spin dynamics in Fig.\ 
\ref{Fig_Tilt_W0}~(b) appears to be even more localized. This may potentially 
be understood due to the additional random disorder $W = 4$ in Fig.\ 4, where 
due 
to rare configurations of the disorder $\epsilon_\ell$ at neighboring sites, 
the difference of 
the neighboring terms $\mu_\ell$ [cf.\ Eq.\ (3) in main text] becomes small 
such that the system behaves more 
ergodic. Moreover, in contrast to our results in Fig.\ 4, we now find that  
${\cal I}_\text{ch}(t)$ and ${\cal 
I}_\text{s}(t)$ in Fig.\ \ref{Fig_Tilt_W0} exhibit pronounced oscillations . 
These ``Bloch oscillations'' are expected in noninteracting models with tilted 
field, but survive to some extent in interacting models as 
well \cite{vanNieuwenburg2019S}.

\section{Using NLCE to study localization dynamics in the 
disordered 
Heisenberg chain}

While we have focused on disordered spin ladders and 
Fermi-Hubbard chains in the main text, the ``standard'' model to study the 
phenomenon of many-body localization is the disordered Heisenberg chain, 
described by the Hamiltonian,
\begin{equation}
 {\cal H}_\text{Heis} = \sum_{\ell = 1}^L {\bf S}_\ell \cdot 
{\bf S}_{\ell+1} + \sum_{\ell = 1}^L h_\ell S_\ell^z\ , 
\end{equation}
where the on-site fields $h_\ell \in [-W,W]$ are drawn at 
random, with $W$ setting the disorder strength.
\begin{figure}[b]
 \centering
 \includegraphics[width=0.95\textwidth]{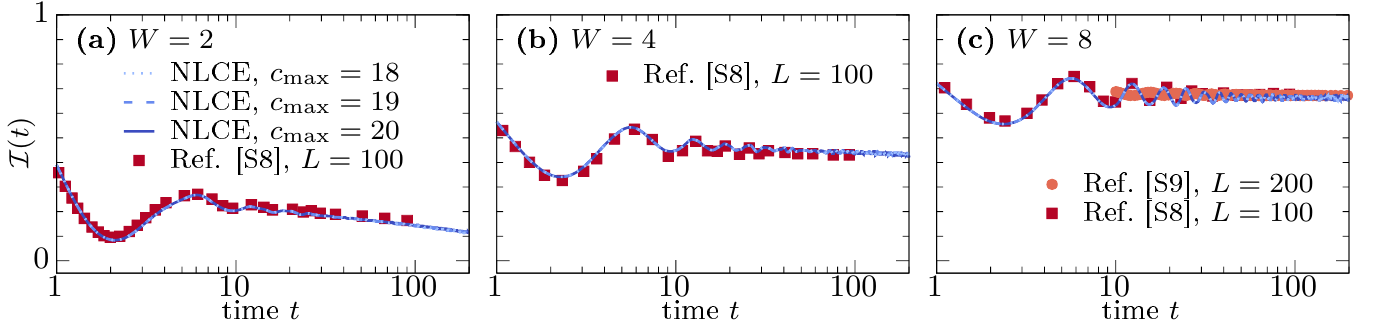}
 \caption{Imbalance ${\cal I}(t)$ in the disordered Heisenberg chain resulting 
from quenches with the antiferromagnetic initial state in Eq.\ 
\eqref{Eq::NeelInit}. NLCE results for expansion orders $c_\text{max} = 
20,19,18$ 
are compared to data obtained by matrix-product-state techniques from Ref.\ 
\cite{Doggen2018S} ($L = 100$) and Ref.\ \cite{Sierant2021S} ($L = 200$). 
Disorder is chosen as {\bf 
(a)} $W = 2$, {\bf (b)} $W = 4$, and {\bf (c)} $W = 8$. Our NLCE data are
averaged over approximately $N_s \approx 2000$ disorder realizations.}
 \label{Fig_Chain}
\end{figure}

It is straightforward to apply the NLCE approach discussed in 
the main part of this paper to study the nonequilibrium dynamics of ${\cal 
H}_\text{Heis}$. To this end, we here focus on the antiferromagnetic initial 
state, 
\begin{equation}\label{Eq::NeelInit}
\ket{\psi(0)} = \ket{\cdots \uparrow \downarrow \uparrow 
\downarrow \cdots}\ , 
\end{equation}
and consider the dynamics of the imbalance ${\cal I}(t) \propto 
\sum_\ell \langle S_\ell^z(t)\rangle \langle S_\ell^z(0)\rangle$.  
Including cluster sizes up to $c_\text{max} \leq 20$ (i.e., the largest 
clusters 
are chains of length $L = 20$ with open boundaries), Figs.\ 
\ref{Fig_Chain}~(a), 
(b), and (c) 
show ${\cal I}(t)$ at $W = 2$, $W = 4$, and $W = 8$, respectively. Considering 
the dynamics of ${\cal I}(t)$ up to times $t \leq 200$, we find that the NLCE 
is well-converged, i.e., the three different expansion orders $c_\text{max} = 
20,19,18$ shown in Fig.\ \ref{Fig_Chain} essentially coincide with each other.  
Moreover, in order to benchmark our NLCE results, we also depict 
in Fig.\ \ref{Fig_Chain} the 
digitized data of Ref.\ \cite{Doggen2018S} 
and Ref.\ \cite{Sierant2021S}, where ${\cal I}(t)$ was obtained using 
matrix-product-state (MPS) techniques. Generally, we find a convincing 
agreement 
between our NLCE results for $L \to 
\infty$ and the MPS data for $L = 100,200$ in the literature.


\begin{thebibliography}{99}
\bibitem{NoteInitState}
For $\ket{\psi_1(0)}$, ${\cal I}_\text{s}(0) = 0$ and we focus on ${\cal 
I}_\text{ch}(t)$ 
in this case.
\bibitem{Note_Error}
We calculate $\Delta {\cal I}$ as 
$\Delta {\cal I} = \frac{1}{\sqrt{N_s}}(\frac{1}{N_s}\sum_s [{\cal I}(t)]_s^2 - 
(\frac{1}{N_s} \sum_s [{\cal I}(t)]_s)^2)^{1/2}$, 
where $[{\cal I}(t)]_s$ denotes the imbalance ${\cal I}(t)$ for a single 
disorder realization and $N_s$ is the number of disorder samples.
\bibitem{Note_NLCE_Psi2}
Since $\ket{\psi_2(0)}$ is in the quarter-filling sector, 
which has a lower Hilbert-space dimension, the computational costs are reduced 
such that we consider slightly larger $c_\text{max}$.
\end{thebibliography}
 \end{document}